\documentclass[conference]{IEEEtran}
\IEEEoverridecommandlockouts

\usepackage{subcaption}
\usepackage{graphicx}
\usepackage{tikz}
\usepackage{float}
\usepackage{dblfloatfix} 
\usepackage{lipsum}
\usepackage{caption}
\usepackage{placeins}
\usepackage{booktabs}
\usepackage{amsmath}
\usepackage{comment}
\usetikzlibrary{arrows.meta, positioning}

\usepackage{fontawesome5} 
\newcommand{\orcid}[1]{\href{https://orcid.org/#1}{\textcolor[HTML]{A6CE39}{$^{\textrm{\faOrcid}}$}}}

\begin{document}

\title{About the Influence of Workflow Topology on Task Intensity
Prediction through Graph Learning}

\author{\IEEEauthorblockN{Max Otto\textsuperscript{1},
Haci Ismail Aslan\textsuperscript{1},
Joel Witzke\textsuperscript{1}, 
Jonathan Bader\textsuperscript{1}, and
Odej Kao\textsuperscript{1}}
\IEEEauthorblockA{\textsuperscript{1}
Dept. Electrical Engineering and Computer Science, Technical University of Berlin, Berlin, Germany}}
\maketitle

\begin{abstract}

Efficient resource provisioning for large-scale workflows on cloud infrastructures is a critical performance engineering challenge. These workflows are often structured as directed acyclic graphs (DAGs), where under-provisioning can cause critical bottlenecks and over-provisioning leads to unnecessary costs. Accurate, task-level prediction of resource intensity (e.g., CPU load and memory usage) is essential for mitigating these issues. While task-level features are commonly used for prediction, the performance impact of the workflow's overall topological structure is often overlooked or assumed. The central question of our work is: To what extent does what part of the DAG topology influence task-level resource intensity, and what is the most effective way to model this influence?

This paper presents a comprehensive benchmark to systematically quantify the impact of graph topology on task intensity prediction. We evaluate and compare a spectrum of modeling approaches. Our findings demonstrate that topology is a critical feature for accurate prediction. Models incorporating important topological information, even through simple handcrafted features, significantly outperform baseline models. We show that graph-native models provide the highest accuracy, achieving low mean absolute errors for both CPU and memory predictions, and can still be combined with simple topological features that they do not learn for better performance.

\end{abstract}


\section{Introduction}
Cloud computing has become the foundation for a wide range of digital services, offering elastic and on-demand access to computational resources~\cite{cloudComputing}. It supports applications in domains such as web search, social networking, e-business, and industrial automation~\cite{cloudComputing, yu2022workflow}. To meet diverse and dynamic demands, modern cloud platforms employ automated resource provisioning and scheduling strategies.

\begin{figure}[h]
    \centering
    \includegraphics[width=\linewidth]{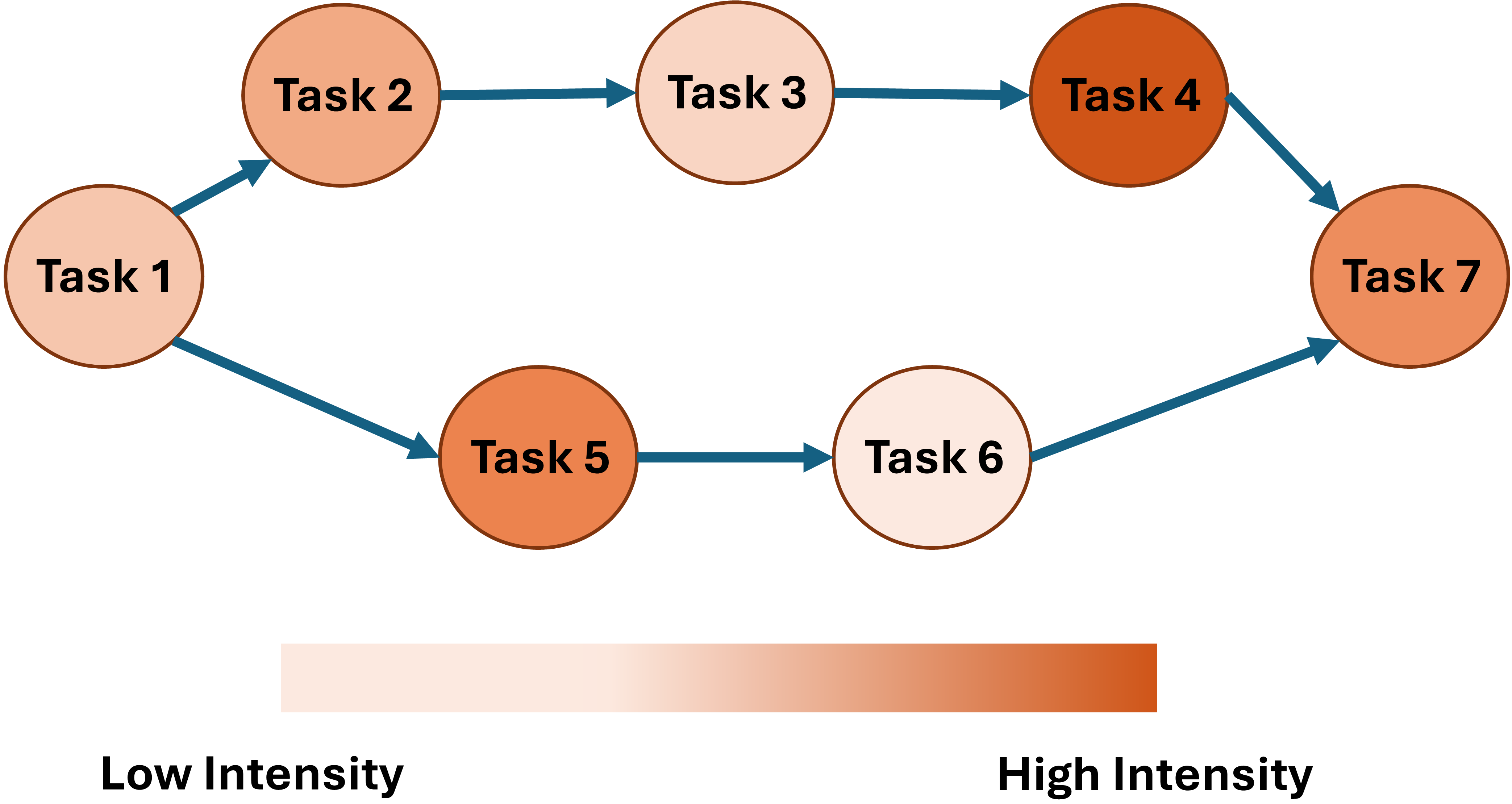}
    \caption{Example workflow DAG with CPU/memory intensities.}
    \label{fig:workflow_max}
\end{figure}

A common abstraction for representing computational processes in the cloud is the workflow, typically modeled as a directed acyclic graph (DAG). In such graphs, nodes represent individual tasks, while directed edges encode dependencies between tasks~\cite{diestel2024graph, ma2021real}, as depicted in Figure \ref{fig:workflow_max}. These dependencies define execution order and often reflect underlying data or control flows in real-world applications.

Each task in a workflow consumes computing resources such as CPU or memory. Accurately estimating these requirements is crucial, especially in shared or pay-per-use environments. Over-provisioning leads to underutilized infrastructure and increased cost~\cite{mehmood2018prediction}, while under-provisioning risks service degradation, bottlenecks, or even task failures~\cite{bader2022leveraging, cong2020survey, masdari2020efficient}. Furthermore, resource demands can exhibit variability and burstiness over time~\cite{ali2014measuring}, making it insufficient to rely solely on historical averages. For effective resource scheduling, systems must account not only for expected usage but also for peaks and anomalies in task intensity.

In recent years, predictive models have been developed to anticipate task-level resource needs prior to execution. Among these, graph-based deep learning methods, particularly Graph Neural Networks (GNNs), have shown promise in capturing both task features and the topological structure of workflows~\cite{yu2022workflow, gao2021workload}. These models exploit the connectivity and hierarchical relationships in workflows to improve predictions of resource usage. However, many existing approaches either target specific model architectures or lack reusability across different scenarios. Furthermore, the relationship between workflow topology and task intensity remains underexplored.

This work systematically studies how the structure of a workflow graph relates to the intensity of its tasks. We approach this in multiple steps. 
First, we train an MLP to predict task intensities on measured task-level features, with and without additional handcrafted topological features, and compare the accuracies. 

Furthermore, we apply various graph learning models for the same task resource predictions, with and without the same topological features as in the MLP experiment, to compare the graph learning models to the MLP and each other, as well as explore the influence of topological features on the models' accuracies.

Finally, we test the real-world applicability of such workflow structure-based approaches. Using the penultimate layer embeddings of a graph learning model that we train on small workflow graphs, in and out of concatenation with topological features, linked with a regressor model, we predict task intensities of larger workflows to explore how such embeddings generalize as well as regression errors.

\section{Related Work}

Prediction of resource consumption is far from a new problem and has already been extensively studied from multiple angles in existing literature.

A fine-grained approach lies in the time series forecasting of resource consumption. Clustering and adaptive network-based fuzzy inference systems (ANFIS)~\cite{5284372}, or an ensemble model using multiple predictor sets with dynamically adjustable members~\cite{cao2014cpu}, yielded promising results for CPU load time series prediction.
Deep learning based on diffusion convolutional recurrent neural networks (DCRNN) was also successfully used for CPU load predictions between 5 and 60 minutes into the future~\cite{al2022deep}.

Within the context of scientific workflows that require high-performance computing, the runtime, disk space, and memory consumption of tasks were estimated using their parameter and input data size~\cite{DASILVA}. To account for estimation errors the authors propose an online estimation process based on the MAPE-K loop (Monitoring, Analysis, Planning, Execution, and Knowledge).
Another approach to predict the task intensity (here, a combination of CPU and memory consumption) has been conducted with reinforcement learning in the form of gradient bandits and evaluated on real-world workflows~\cite{bader2022leveraging}.
Memory predictions of tasks also enable adjustable resource scheduling models for scientific workflows, which can either focus on high throughput or minimization of resource usage~\cite{8066333}. The models are based on the slow-peaks model and thus assume resource exhaustion towards the end of a task's execution time.

Previous work also investigated the use of graph features and GNNs.
One implements a scheduling framework that combines graph neural networks and deep reinforcement learning (DRL) using a reinforcement learning agent that is trained through the proximal policy optimization (PPO) algorithm to optimize makespan and energy consumption in dynamic cloud workflow environments~\cite{chandrasiri2025energy}.
A different approach using various graph neural network models on the offline cloud workflow database Alibaba Cluster-Trace-V2018 concluded that homogeneous graphs are a better base for GNNs, than heterogeneous graphs~\cite{gao2021workload}. Furthermore, on the dataset, graph attention networks (GAT) are slightly inferior to graph convolutional networks (GCN), and GNNs predict task performance more accurately for tasks located later in the workflow.

The same dataset was also used for task intensity prediction with a transformer model that encodes the DAG information in a position embedding block~\cite{yu2022workflow}. Experiments showed that the transformer model received better results with graph structure information than without. Also, two graph topology-based models are among the best three evaluated machine learning models of the paper.

The last two approaches, which include graph learning for task resource estimation, indicate a connection between task resource intensity and topological graph or task information, which is explored, in detail and in connection to particular structural components, throughout this paper, within the scope of static offline cloud computing workflows and graph-based deep learning. 

\section{Methods}

\subsection{About the Applied Deep Learning Models}
\label{sec:AboutModels}
The following sections provide information about the different deep learning models that are applied throughout this paper. The majority of these models leverage topological information in order to infer task intensities through different approaches. Each model exhibits distinct strengths.
\subsubsection{Multi Layer Perceptron (MLP)}
The multi-layer perceptron \cite{popescu2009multilayer} \cite{noriega2005multilayer} is a widely used and very basic neural network, that does not apply any topological techniques by itself and is used as a baseline network for this paper.

\subsubsection{Graph Convolutional Networks (GCN)}
Graph convolutional networks \cite{kipf2016semi} are neural networks of the GNN family, based on convolutional neural networks. They are explicitly designed for applications where the data is encoded in the form of graphs and are based on message passing.

\subsubsection{Graph Attention Networks (GATs)}
Graph attention networks \cite{velivckovic2017graph} are graph neural networks that bring an attention mechanism into the neighborhood message-passing.

\subsubsection{Graph Isomorphism Networks (GINs)}
Graph isomorphism networks \cite{xu2018powerful} are GNNs that achieve the same discriminative power as the  and the Weisfeiler-Lehman graph isomorphism test \cite{weisfeiler1968reduction}. 

\subsubsection{GraphSAGE}
GraphSAGE (SAmple and aggreGatE) \cite{hamilton2017inductive} is a GNN that aggregates neighbor features in a $k$-hop fashion. It starts by aggregating immediate neighbors and iteratively increases the search depth $k$ to receive information from more distant neighborhoods. This mechanism allows GraphSAGE to capture both local and higher-order graph structures efficiently.

\subsubsection{DAGTransformer}
\label{sec:DAGTransformer}
The DAGTransformer, introduced in \cite{yu2022workflow}, is a transformer encoder model that leverages task positions within the graph as positional encoding. Moreover, it uses the neighbors of tasks as an attention mask, enabling a GNN-like aggregation mechanism.

\subsubsection{H2GCN}
\label{sec:h2gcnexplained}
H2GCN \cite{zhu2020beyond} is a graph neural network, that was designed to tackle the challenges that are raised by heterophile graphs (graphs in which neighboring nodes often have very dissimilar labels), by employing special aggregation and embedding strategies.

\subsubsection{Ensemble Models}
\label{sec:ensemble}
Ensembles combine multiple weaker classifier models into a more robust and more efficient one \cite{manconi2022soft}. For its simplicity and good performance, this paper utilizes supervised soft voting \cite{manconi2022soft} ensembles of combined graph learning networks.

\subsubsection{XGBoost}
XGBoost \cite{chen2016xgboost} is a scalable system that handles nonlinearity by using a boosted tree structure, instead of activation functions as neural networks do. We utilize XGBoost for downstream regression experiments based on penultimate layers of pre-trained classification networks.

\subsection{Utilized Topological Features}
\label{sec:topofeat}
To evaluate the impact of network topology in our experiments, we need a set of corresponding topological features. Many such features can be derived directly from the graph's structure. The following sections describe these features in detail. 
\subsubsection{Node-Based Features}
\label{sec:node_based_features}
\paragraph{Node Path Length.} In a graph, the path length between two nodes is the minimum number of edges that connect them. For this paper, we construct a feature vector for each node by calculating its path length to every other node in the graph. For example, a feature we call \emph{node 1 path length} refers to the path length from the target node (i.e., the one being predicted) to `task node 1' of the workflow.

\paragraph{DAG In \& Out.} Input and output tasks of the current task are specified in the form of a flattened sparse matrix called \emph{DAG in \& out}~\cite{yu2022workflow}. In simpler terms, this is the normalized adjacency matrix for the given task. For a workflow with 7 tasks, the DAG in \& out feature vector for each task has 15 dimensions. These dimensions are composed of 7 values for incoming edges, 7 for outgoing edges, and one final value for the task's node ID.

\paragraph{Node Degree.} Node degree  describes the number of edges of a node~\cite{graphtheorybionet}, treating each neighboring node equally.

\paragraph{Node Centrality.} Node centrality emphasizes the importance of nodes in a graph. Many centralities exist, and in this paper, three of them are utilized. These are eigenvector centrality, betweenness centrality, and closeness centrality:

Eigenvector centrality~\cite{ruhnau2000eigenvector} is a score that is  based on the premise, that a node in a graph is important if the node's neighbors are important.

By the intuition behind betweenness centrality~\cite{barthelemy2004betweenness}, a node is important, if it lies on many shortest paths between other nodes in the graph.

Closeness centrality~\cite{okamoto2008ranking} measures the lengths of paths between the node $v$ and other nodes: If a node is part of many shortest path lengths to other nodes in a graph, it is well connected and therefore important.

\paragraph{Clustering Coefficient.} \label{sec:clustCoeff}
The local clustering coefficient measures how connected a node's neighbors are within the graph~\cite{li2017clustering, soffer2005network}.

\paragraph{Node2Vec.} \label{sec:Node2Vec}
Node2Vec~\cite{grover2016node2vec} is an unsupervised method of learning node embeddings in such a way, that nodes from the same neighborhood receive similar embeddings, based on random walks.

\paragraph{Color Refinement and Weisfeiler-Lehman (WL) Graph Kernel.} \label{sec:WL}
The color refinement (or naive vertex refinement) algorithm is a graph isomorphism test, that iteratively refines node colorings based on structural similarities.

During each iteration, the color of a node is updated based on its current color and the colors of its neighbors~\cite{douglas2011weisfeiler} and compressed into a new color.
We use the final colors from the color refinement algorithm directly as node features.

\paragraph{Graphlets.}
Subgraph patterns within a larger graph are called graphlets~\cite{ahmed2015efficient}. They represent the local structure of a graph by considering the connectivity patterns of a fixed number of nodes and edges.

This paper counts the occurrence of the task node in graphlets with up to three task nodes (edges, paths, triangles). The number of occurrences in each graphlet type is then passed as a feature embedding vector.

\subsubsection{Katz Index}
\label{sec:Katz}
The Katz index~\cite{zhan2017identification} is a method used to measure the influence of a node within a graph, taking into account both direct and indirect connections. Each row of the Katz matrix captures the influence of a specific node on all other nodes in the graph with respect to direct and indirect connections. These influences are aggregated, resulting in a vector, that stores the total influence of each node across the entire graph. This vector can be used as a node feature vector.

Moreover, by aggregating the entries of the final influence vector, a graph-level Katz index can be acquired.

\subsection{Evaluation Metrics} Evaluating a deep learning model requires selecting appropriate metrics that reflect its performance on a given task. This paper mainly uses accuracy \cite{sokolova2006beyond}. It is a common metric for classification problems, measuring the proportion of correctly predicted samples. 

To evaluate regression models, we utilize Mean Absolute Error (MAE) \cite{qi2020mean}, which describes the average size of errors within the set of predictions, as well as the coefficient of determination (or $R^2$, R 2) \cite{zhang2017coefficient}, which is a value between 0 and 1 that explains the variation in the model's predictions. A model that always predicts the average has an R-squared value of 0, and a perfectly fitting model has an R-squared value of 1.

\subsection{Alibaba Cluster-Trace-V2018 Dataset}
\label{sec:dataset}
Our experiments are performed on the Alibaba Cluster-Trace-V2018 based dataset from Yu et al.\cite{yu2022workflow}. It consists of cloud workflows that include 7 tasks, and the CPU and memory usages of the 7th tasks are predicted. These workflows have been extracted from two $.csv$ files, $batch\_task$ and $batch\_instance$, which can be found within the Cluster-Trace-V2018 Github.

For each task, the authors extract 34 measured task-level features, mostly by aggregating instance-level features, such as CPU and memory usages or execution times. They create class labels by clustering mean and maximum of the task's according usages.

The dataset is split into multiple splits, differing by training set size, validation set size, and test set size.

For example, when applying the $split6\_2\_2\ $, Yu et al. utilize 60\% of the data for training, and 20\% for validation and testing, respectively.

\section{Determining Topological Feature Importances with MLP}
\label{sec:MLPTOP}
This section details two experiments designed to evaluate the importance of different topological features when predicting task resource usage. To achieve this, we train a Multi-Layer Perceptron (MLP) on workflows from the Alibaba dataset (see Section \ref{sec:dataset}) and compare the model's performance using different topological feature combinations.

\paragraph{First MLP Experiment.}
Within the scope of the first experiment, the MLP only learns from topological features of the predicted task: the only inputs that the MLP learns from are the different structural features that are compared and combined, including a baseline feature of node lengths to 7th tasks (with a variance of 0). Thus, the input vector for the target task (e.g., the 7th task in a workflow) contains a set of topological features of only the predicted 7th task.

\paragraph{Results of First MLP Experiment.}
The results of the first experiment are depicted in Table \ref{tab:MLP}, demonstrating MLP classification accuracies of the mentioned 7th tasks with varying topological input features over various splits\footnote{A\ 0.6/0.2/0.2\ split (1\textsuperscript{st} split) for\ memory\ and\ 0.6/0.2/0.2,\ 0.8/0.1/0.1\ (2\textsuperscript{nd} split) and\ 0.9/0.05/0.05\ (3\textsuperscript{rd} split) for\ CPU, where the first, second, and third numbers represent the percentages for the training, validation, and test sets, respectively.} of the mentioned dataset. For CPU intensity prediction, large increases in accuracy are achieved by introducing Weisfeiler-Lehman-based features, as described in \ref{sec:node_based_features}. Katz indices and the centrality features (concatenated into `3 centralities'), as well as node path lengths, which are all explained in Section \ref{sec:topofeat}, also enhance the prediction accuracies of the MLP significantly. It is noteworthy, that path lengths to tasks further away are superior to closer tasks. This is likely due to more variance in the 7th tasks' path lengths to further tasks.

Node degrees, clustering coefficients and graphlets perform very similarly, likely because these three features share many characteristics.

The findings of the first experiment for the MLP in memory intensity prediction also emphasize the importance of topological information in predicting the memory intensity. The MLP's base line accuracy for memory prediction from 7th tasks node path lengths starts at 90.98\%.

While only a few structure-based features improve the memory accuracy over 90.98\% alone, when introducing all workflow structure features to the MLP together, they interact in a way that increases accuracy by almost 4\%. The highest accuracy is gained when combining all topological features during CPU prediction as well, stressing the importance of this feature interaction.

The ineffectiveness of Node2Vec based features (Section \ref{sec:Node2Vec}) might be explained by the small graph size of 7 task workflows. With the small graph size, there could not be enough variance in task neighborhoods to form expressive embeddings.

In summary, the first MLP experiment shows that adding topological information of the predicted tasks leads to higher accuracy than the baseline features, indicating that the tasks' structural characteristics alone provide useful information for predicting task intensity, when using a very basic neural network. 

\begin{table*}[h]
\centering
\begin{tabular}{lcccc} 
\toprule
Feature Method & CPU Accuracy (1\textsuperscript{st} Split) & CPU Accuracy (2\textsuperscript{nd} Split) & CPU Accuracy (3\textsuperscript{rd} Split) & Memory Accuracy \\ 
\midrule
DAG In \& Out & 51.87\% & 52.18\% & 51.99\% & 91.65\%\\
Node Degrees & 48.27\% & 48.24\% & 48.15\% & 90.98\%\\
Clustering Coefficient  & 48.14\% & 48.25\% & 48.18\% & 90.98\%\\
Graphlets & 48.27\% & 48.10\% & 48.39\% & 90.98\%\\
Node 1 Path Length& 70.74\% & 70.50\% & 70.00\% & 90.98\%\\
Node 2 Path Length & 65.55\% & 63.30\% & 64.58\% & 90.98\% \\
Node 3 Path Length & 54.93\% & 54.03\% & 55.24\% & 90.98\%\\
Node 4 Path Length & 47.16\% & 46.41\% & 47.30\% & 90.98\%\\
Node 5 Path Length & 45.53\% & 45.39\% & 44.37\% & 90.98\%\\
Node 6 Path Length & 44.55\% & 45.47\% & 44.99\% & 90.98\%\\
Node 7 Path Length & 45.34\% & 45.46\% & 44.67\% & 90.98\%\\
All Path Lengths & 74.56\% & 74.26\% & 73.88\% & 93.06\%\\
Weisfeiler-Lehman & 69.70\% & 70.15\% & 68.24\% & 90.98\%\\
Katz Nodelevel & 62.05\% & 59.19\% & 62.33\% & 90.98\%\\
Katz Graphlevel & 65.55\% & 65.91\% & 65.76\% & 90.98\%\\
Eigenvector Centrality & 63.61\% & 64.74\% & 73.88\% & 90.98\%\\
Betweenness Centrality & 47.75\% & 47.83\% & 48.36\% & 91.69\%\\
Closeness Centrality & 66.62\% & 67.05\% & 66.61\% & 90.98\%\\
3 Centralities & 73.97\% & 74.12\% & 73.44\% & 92.89\%\\
Node2Vec embedding & 45.15\% & 44.79\% & 44.90\% & 90.98\%\\
All Topology Features & 75.93\% & 75.76\% & 75.07\% & 94.42\%\\
\bottomrule
\end{tabular}

\caption{Impact of topological features on CPU and memory intensity prediction (experiment 1).}
\label{tab:MLP}
\end{table*}

\paragraph{Second MLP Experiment.}
Within the scope of the second experiment, the MLP learns task intensities from the entire workflow instead of only the predicted task, including the following features:
\begin{itemize}
    \item the measured task level intensity features of all earlier tasks, explained in Section \ref{sec:dataset}
    \item topological features of all earlier tasks and the predicted task
\end{itemize}
This approach allows for the exploration of how each topological feature, in combination with measured features and other tasks' topological features, influences accuracy, as opposed to only leveraging the predicted task's topological features, as in the first experiment, creating a link from MLP to graph learning. 
\begin{table*}[h]
\centering
\begin{tabular}{lcccc} 
\toprule
Feature Method & CPU Accuracy (1\textsuperscript{st} Split) & CPU Accuracy (2\textsuperscript{nd} Split) & CPU Accuracy (3\textsuperscript{rd} Split) & Memory Accuracy \\ 
\midrule
Only Measured Features & 86.60\% & 86.82\% & 87.00\% & 96.21\%\\
Only All Topology Features & 77.63\% & 77.20\% & 77.55\% & 95.04\%\\
DAG In \& Out & 90.16\% & 89.43\% & 90.22\% & 98.51\%\\ 
Clustering Coefficient & 88.59\% & 87.53\% & 87.83\% & 97.80\%\\
Node Degrees & 88.33\% & 87.40\% & 88.86\% & 97.72\%\\
Graphlets & 87.98\% & 87.66\% & 89.25\% & 97.76\% \\
All Path Lengths & 88.91\% & 87.65\% & 88.86\% & 97.73\%\\
3 Centralities & 89.04\% & 88.28\% & 89.36\% & 97.87\%\\
Weisfeiler-Lehman & 88.38\% & 87.84\% & 89.01\% & 97.31\%\\
Katz Nodelevel & 89.30\% & 87.29\% & 88.50\% & 97.89\%\\
All Topology Features & 89.15\% & 89.00\% & 89.71\% & 98.21\%\\
\bottomrule
\end{tabular}

\caption{Impact of topological features when topology information is considered for all nodes (experiment 2).}
\label{tab:MLP_measured}
\end{table*}

\paragraph{Results of Second MLP Experiment.} The results of the second experiment are depicted in Table \ref{tab:MLP_measured}, which is constructed similarly to the table from the first MLP experiment. It shows that the MLP benefits from the inclusion of previous task's topological features during training without measured features, in comparison to the accuracies of the MLP in experiment 1. However, the difference here is not immense, so the topology of the predicted task's node is likely the most important. In combination with measured features, all implemented topological features add to the performance of the MLP that is trained on measured features, displaying the effectiveness of the combination of topological and measured features of all tasks.

Similarly to the first experiment, for memory and CPU intensity prediction, the MLP can take great advantage of Weisfeiler-Lehman and centrality-based features, but also topological features that were not as promising in experiment 1. The most dominant topological feature in experiment 2 is the \emph{DAG in\&out} sparse matrix, which contains the normalized adjacency list. Since it holds the entire workflow structure now, this is expected.

Moreover, all other employed structural features increase the MLP accuracy significantly into a similar range of accuracy. Additionally, training the MLP using a combination of all topological features also increases accuracy drastically, only surpassed by the sparse matrix.

The results of the second MLP experiment demonstrate that incorporating the topological features of all tasks within the workflow is highly beneficial for task-intensity prediction using a basic deep learning model, substantially improving the accuracy of the MLP even when only measured features are used.

\section{Graph Deep Learning on Workflows}

\label{sec:modelsnewds}

\begin{table*}[h]
\centering

\begin{tabular}{lccc}
\toprule
GNN &  CPU Acc. (1\textsuperscript{st} Split)  & CPU Acc. (2\textsuperscript{nd} Split) & CPU Acc. (3\textsuperscript{rd} Split)\\
\midrule
GCN & $88.38 \pm 0.4\%$ & $88.18 \pm 0.2\%$ & $88.91 \pm 0.1\%$\\
GAT & $89.81 \pm 0.2\%$ & $89.61 \pm 0.2\%$ & $90.48 \pm 0.3\%$\\
GIN & $89.87 \pm 0.1\%$ & $89.63 \pm 0.2\%$ & $90.64 \pm 0.3\%$\\
GraphSAGE & $87.80 \pm 0.2\%$ & $87.35 \pm 0.2\%$ & $88.18 \pm 0.2\%$\\
H2GCN & $87.84 \pm 0.1\%$ & $87.11 \pm 0.1\%$ & $88.20 \pm 0.1\%$\\
DAGTransformer (stated in paper) & $91.25 \pm 0.04\%$ & $91.11 \pm 0.05\%$ & $92.15 \pm 0.13\%$\\
DAGTransformer (reproduced) & $90.54 \pm 0.1\%$ & $89.96 \pm 0.05\%$ & $90.61 \pm 0.3\%$\\
GNN Ensemble & $91.12 \pm 0.1\%$ & $90.52 \pm 0.2\%$ & $91.59 \pm 0.2\%$\\
\bottomrule
\end{tabular}

\caption{Graph learning models CPU average prediction accuracies without topological features.}
\label{tab:GNNs CPU}
\end{table*}

\begin{table*}[t]
\centering

\begin{tabular}{lccc}
\toprule
GNN &  CPU Acc. (1\textsuperscript{st} Split) & CPU Acc. (2\textsuperscript{nd} Split) & CPU Acc. (3\textsuperscript{rd} Split)\\
\midrule
GCN & $90.34 \pm 0.3\%$ & $89.67 \pm 0.1\%$ & $90.35 \pm 0.4\%$ \\
GAT & $90.54 \pm 0.1\%$ & $89.89 \pm 0.1\%$ & $90.89 \pm 0.2\%$ \\
GIN & $90.56 \pm 0.1\%$ & $90.14 \pm 0.2\%$ & $90.74 \pm 0.05\%$ \\
GraphSAGE & $88.97 \pm 0.4\%$ & $88.08 \pm 0.3\%$ & $89.10 \pm 0.1\%$ \\
H2GCN & $89.31 \pm 0.1\%$ & $88.73 \pm 0.3\%$ & $89.81 \pm 0.2\%$ \\
DAGTransformer (stated in paper) & - & - & -\\
DAGTransformer (reproduced) & $90.75 \pm 0.1\%$ & $90.37 \pm 0.2\%$ & $90.89 \pm 0.2\%$ \\
GNN Ensemble & $91.21 \pm 0.1\%$ & $90.24 \pm 0.1\%$ & $91.45 \pm 0.3\%$ \\
\bottomrule
\end{tabular}

\caption{Graph learning models CPU average prediction accuracies with topological features.}
\label{tab:GNNs CPU topo}
\end{table*}

The experiments in Section \ref{sec:MLPTOP} established that workflow topology is a useful feature for predicting task intensities. Building on these findings, this section transitions to the evaluation of the same dataset with graph-based learning models, which are specifically designed for such data.

 The models that we decided to compare to each other, and to the MLP of the previous section, include GIN, GAT, GraphSAGE, as well as an ensemble of these three. Moreover, we benchmark the performance of H2GCN, GCN and DAGTransformer. All models are explained in Section \ref{sec:AboutModels}.
All models are trained on the 7th task of each workflow, with and without all of the additional structural features for all tasks, that were employed in section \ref{sec:MLPTOP}, to explore a possible impact of additional topological information on the graph learning results. The accuracies of these models on the Alibaba Cluster-Trace-V2018 dataset can be found in Tables \ref{tab:GNNs CPU} and \ref{tab:GNNs CPU topo} for CPU usage prediction and Table \ref{tab:GNNs mem} for memory usage prediction. In the tables, \texttt{acc.} refers to accuracy.

\paragraph{Discussion of the Graph Learning Results.}
We find that the GNN accuracies lay in a similar area to the MLP accuracies of experiment 2 in Section \ref{sec:MLPTOP}. While GCN, GraphSAGE, and H2GCN perform slightly worse in comparison to the MLP with all nodes' topological information, GAT, GIN, DAGTransformer, and the GNN ensemble outperform it, even without the addition of features that describe the workflow structures.

\begin{table*}[h!]
\centering

\begin{tabular}{lcc} 
\toprule
GNN & Memory Accuracy & Memory Accuracy with Topological Features\\ 
\midrule
GCN & $97.67\pm0.03\%$ & $98.52\pm0.03\%$ \\
GAT & $98.34\pm0.1\%$ & $98.68\pm0.1\%$ \\
GIN & $98.47\pm0.2\%$ & $98.59\pm0.1\%$ \\
GraphSAGE & $97.33\pm0.1\%$ & $97.78\pm0.4\%$ \\
H2GCN & $90.98\pm0.01\%$ & $90.98\pm0.004\%$ \\
DAGTransformer (stated) & $98.56\%$ & - \\
DAGTransformer (reproduced) & $98.35\pm0.1\%$ & $98.59\pm0.1\%$ \\
GNN Ensemble & $98.32\pm0.2\%$ & $98.36\pm0.3\%$ \\
\bottomrule
\end{tabular}

\caption{Graph learning models memory average prediction.}
\label{tab:GNNs mem}
\end{table*}

The strong performance of the GIN model implies that workflow isomorphism is an important factor in predicting CPU and memory intensities. This finding is consistent with the accuracy improvements observed in Section \ref{sec:MLPTOP}, where introducing Weisfeiler-Lehman labels to the MLP also enhanced its performance. This significant influence of isomorphism may be attributed to the dataset's composition, as it consists exclusively of 7-task workflows. Such uniformity likely leads to a high number of isomorphic graphs, causing certain structural patterns to repeat frequently within the data.

The efficiency of GAT can possibly be traced back to its attention mechanism, which enables it to recognize more complex patterns than other GNN networks. It allows the GAT network to learn weights for a task's neighbors. These neighbor weights can be learned by patterns such as the neighbor's influence on other nodes, which ultimately defines node centrality as described in Section \ref{sec:node_based_features}. Since leveraging task centralities as well as the Katz index helped the MLP network in Section \ref{sec:MLPTOP} to enhance its accuracy when predicting task intensities a lot, these could be the key patterns for the GAT performances.

Although GraphSAGE alone does not perform as well as other models, it enhances the performance of the GNN ensembles; thus, its special aggregation strategy might add to the features of the other mentioned GNNs.

From the H2GCN accuracies, it can be deducted that workflow heterophily is not very problematic on the dataset, since other GNNs were
able to outperform the H2GCN, which was explicitly created to tackle graphs with strong
heterophilic properties that negatively affect other GNNs.

Lastly, while we were not able to exactly reproduce the claimed accuracies, the DAGTransformer performed extremely well, which is expected, as the dataset was also used in the original DAGTransformer paper. It should be mentioned that the DAGTransformer also attends to different neighboring nodes in different ways, so it likely learns similar patterns as the GAT. 

Furthermore, by comparing the model performances in Tables \ref{tab:GNNs CPU}, \ref{tab:GNNs CPU topo}, and \ref{tab:GNNs mem}, we find that the accuracies of almost all introduced geometric deep learning models are slightly improved when additional graph structural features are included, compared to when these features are not present, allowing GCN and partially H2GCN to exceed the MLP accuracies.
This leads to the conclusion that even geometric neural networks do not have the ability to capture some important graph structure of the cloud workflows, which can be leveraged for better intensity predictions. 

The only exception to this behavior is the GNN ensemble, which covers a wide area of graph structure by evaluating multiple graph-based networks. From this behavior, we can derive that it already gathers a large part of the information that the additional topology-based features introduce to other models by itself.

\section{Transferability of Embeddings: Regressing the Actual CPU \& Memory Utilization}
\label{cha:regressions}

Previously, throughout this paper, all employed deep learning and graph learning models were trained and evaluated on 7-task large workflow graphs, and using classification accuracy. While classification is helpful for benchmarking models in this case, it might not represent use cases of such graph learning models in the real world. 

For real applications, it is most likely that predictions would need to be made over various workflow graph sizes, and exact predictions might be more useful than classification. Therefore, a regression model would be more applicable.

Moreover, it would not be computationally efficient to train a large graph learning model for each workflow size. While there are likely many possible pipelines to approach this problem, a good alternative that fits our structure could be inferring the penultimate layer of a pre-trained model, which contains aggregated structure information (if it can generalize), on new graph sizes, and train a small regressor on that layer to transfer the model information for each workflow size. 

\subsection{The Regression Experiment}
To simulate such an approach, we introduce a dataset of 20th and 18th tasks, in order to test the graph learning models' generalization capabilities. It is extracted from the Alibaba Cluster-Trace-V2018 dataset similarly as explained in section \ref{sec:dataset}, except for workflow size. Instead of clustered class labels, we use measured data of the predicted tasks as ground truth. 

By inferring with pre-trained models and training on the resulting penultimate layers, we explore CPU and memory regression, with and without using the topological features.

Since it is the best performing model that profits from the topological features that we benchmarked for classification, we leverage the pre-trained DAGTransformer embeddings to train an XGBoost regressor model, that predicts CPU and memory usage of the 18th and 20th tasks.

First, the XGBoost model is fed by penultimate layer embeddings of the standard DAGTransformer.
Afterwards, all previously introduced topological features are computed for the predicted tasks and aggregated with the penultimate layer embeddings to explore a possible enhancement of the XGBoost's predictions. 

\subsection{Results of the XGBoost Model on Penultimate Layers}

\begin{figure}[!h]
    \centering
    \begin{subfigure}[h]{0.5\textwidth}
    \includegraphics[width=\linewidth, height=4cm]{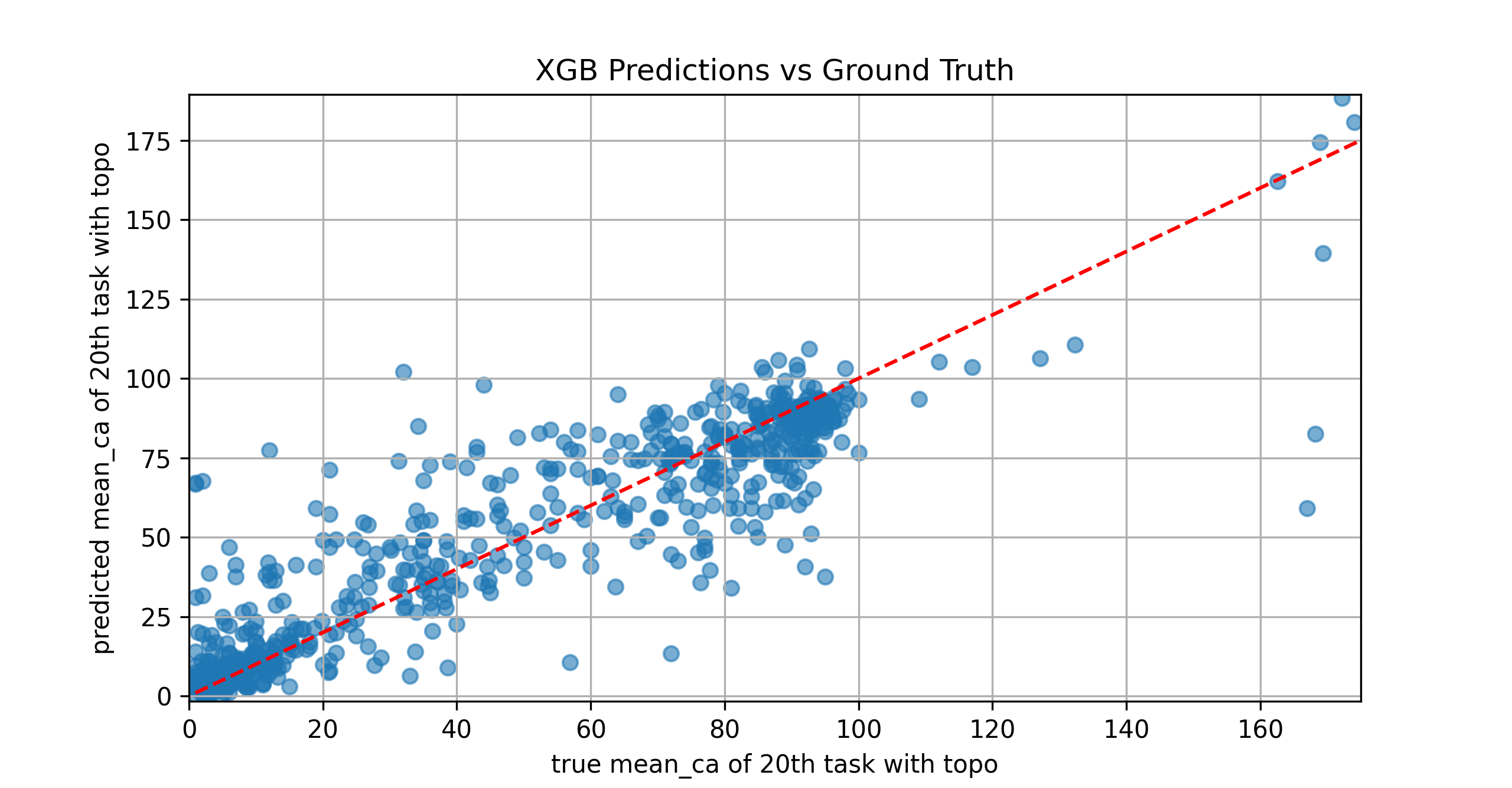}
    \caption{Regression of measured instance CPU average aggregated by mean for 20th tasks with topology.}
    \label{fig:20_mean_cm_topo}
    \end{subfigure}

    \centering
    \begin{subfigure}[h]{0.5\textwidth}
    \includegraphics[width=\linewidth, height=4cm]{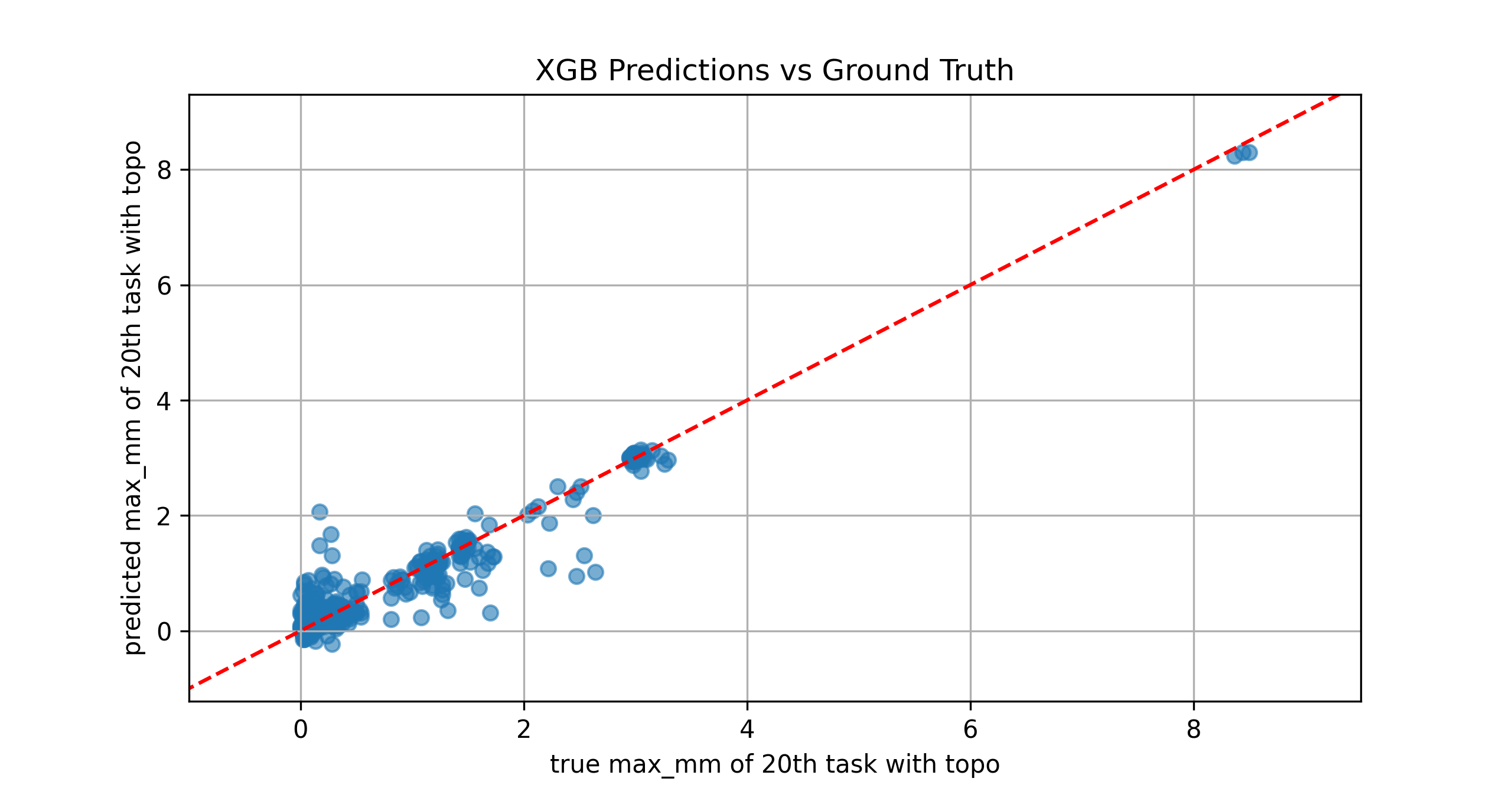}
    \caption{Regression of measured instance memory maximum aggregated by maximum for 20th tasks with topology.}
    \label{fig:20_max_mm_topo}
    \end{subfigure}
    \caption{Regression plots when model embeddings are used to feed XGBoost.}
    \label{fig:regression_full}
\end{figure}

Tables \ref{tab:MAE} and \ref{tab:r2} show mean absolute error and $R^2$ scores of these XGBoost models: \texttt{avg} is short for average, \texttt{max} for maximum, and \texttt{Topo} references additional topological features of the predicted task.
These results, as well as the regression plots in Figures \ref{fig:20_mean_cm_topo} to \ref{fig:20_max_mm_topo}, show that the penultimate layers can be useful for building regression models that predict task-level CPU and memory utilization, even for larger workflows, while weaker for most \texttt{max CPU} predictions.

\begin{table}[h]
    \centering
        \begin{tabular}{lcccc}
        \toprule
        task  & avg CPU & max CPU & avg mem & max mem \\
        \midrule
        20 Mean & 8.42 & 22.76 & 0.085 & 0.10 \\
        20 Mean Topo & 7.77 & 22.51 & 0.074 & 0.09 \\
        20 Max & 9.79 & 34.60 & 0.11 & 0.12 \\
        20 Max Topo & 9.53 & 33.11 & 0.10 & 0.10 \\
        18 Mean & 11.51 & 30.24 & 0.06 & 0.08 \\
        18 Mean Topo & 10.67 & 28.42 & 0.06 & 0.08 \\
        18 Max & 13.40 & 52.44 & 0.08 & 0.09 \\
        18 Max Topo & 12.75 & 52.80 & 0.08 & 0.09 \\
        \bottomrule
    \end{tabular}
    \caption{MAE values for average CPU \& memory utilization.}
    \label{tab:MAE}
\end{table}

\begin{table}[h]
    \centering
        \begin{tabular}{lcccc}
            \toprule
            task  & avg CPU & max CPU & avg mem & max mem \\
            \midrule
            20 Mean & 0.85 & 0.41 & 0.95 & 0.94 \\
            20 Mean Topo & 0.87 & 0.51 & 0.96 & 0.95 \\
            20 Max & 0.83 & 0.45 & 0.93 & 0.92 \\
            20 Max Topo & 0.84 & 0.50 & 0.94 & 0.94 \\
            18 Mean & 0.79 & 0.68 & 0.41 & 0.61 \\
            18 Mean Topo & 0.81 & 0.75 & 0.46 & 0.60\\
            18 Max & 0.77 & 0.66 & 0.53 & 0.59 \\
            18 Max Topo & 0.80 & 0.64 & 0.39 & 0.67 \\
            \bottomrule
        \end{tabular}
        \caption{$\text{R}^2$ scores for average CPU \& memory utilization.}
        \label{tab:r2}
\end{table}

The Tables \ref{tab:MAE} and \ref{tab:r2} show, that mean absolute errors and $R^2$ of regression predictions of 20th and 18th tasks reach MAEs down to $8.42$ for CPU prediction and $0.085$ for memory prediction. The regression plots in Figure \ref{fig:regression_full} show that wrong predictions in task intensities are often those with very high values, which could be traced back to a low amount of such training data within the utilized dataset, but generally, the predictions are close to the diagonal.

Moreover, as shown in Tables \ref{tab:MAE} and \ref{tab:r2}, training the regression model with a concatenation of the penultimate layer and additional topological information of the predicted task node as input features increases its performance, even though the DAGTransformer already collects topological information. This reinforces that even graph-learning-based embeddings/models cannot capture the entire topological image of the predicted task/workflow.

\section{Conclusion}
\label{cha:conclusion}

The results of this paper show that learnable patterns connect many different aspects of workflow topology and task intensity.

Thus, by introducing different aspects of task and workflow topology as features, even simple models can efficiently explore a variety of patterns, and by combining these features, a high accuracy can be achieved, as shown by the experiments in Section \ref{sec:MLPTOP}.

Moreover, we explored a variety of graph-learning-based models in Section \ref{sec:modelsnewds}, which mostly outperformed the MLP. 

While graph learning is very effective at predicting CPU and memory usage of future tasks, even these models do not always capture all aspects of topology that are useful. Thus, topology-based features can also slightly improve the performance of graph learning models. 

Next, we explored the generalization of Graph learning and topological features by combining a DAGTransformer backbone with an XGBoost regressor. While trained on 7-task graphs, the model effectively generalizes to 18 and 20-task workflows, but additional topological input features furthermore enhanced the MAE's of the pipeline.

In essence, the results of this paper show that it can be advantageous to add many different aspects of workflow graph structure to a model's input features when predicting task CPU and memory needs, even if the model already collects some graph information.

\section*{Acknowledgements}

This paper was supported by the Swarmchestrate project of the European Union’s Horizon 2023 Research and Innovation programme under grant agreement no. 101135012 and the Deutsche Forschungsgemeinschaft (DFG, German Research Foundation) as FONDA (Project 414984028, SFB 1404).
\bibliographystyle{IEEEtranS} 
\bibliography{refs} 

\newpage

\FloatBarrier
\end{document}